\documentclass[twocolumn]{aastex62} 
\usepackage{amsmath}
\usepackage{longtable, booktabs, threeparttablex}

\graphicspath{{./}{figures/}}

\shorttitle{Limits on Stellar Flybys in the Solar Birth Cluster}
\shortauthors{Siraj, Chyba, \& Tremaine}

\begin{document}

\title{Limits on Stellar Flybys in the Solar Birth Cluster}

\email{siraj@princeton.edu}

\author{Amir Siraj}
\affil{Department of Astrophysical Sciences, Princeton University, 4 Ivy Lane, Princeton, NJ 08544, USA}

\author{Christopher F. Chyba}
\affil{Department of Astrophysical Sciences, Princeton University, 4 Ivy Lane, Princeton, NJ 08544, USA}
\affil{School of Public and International Affairs, Princeton University, 20 Prospect Lane, Princeton, NJ 08540, USA}

\author{Scott Tremaine}
\affil{School of Natural Sciences, Institute for Advanced Study, Princeton, NJ 08540, USA}

\begin{abstract}

The orbits of small bodies in the outer solar system are particularly sensitive to gravitational perturbations, including stellar flybys. Stellar clusters, with low velocity dispersions and high number densities, can be the source of strong and frequent flybys. As a result, we can infer what properties of the solar birth environment would be incompatible with the structure of the outer solar system observed today. Here, we explore with $n-$body simulations the implications of the low inclinations ($i < 20^{\circ}$) of the distant sednoids (objects with perihelia $q > 40 \mathrm{\; AU}$ and semimajor axes $a > 400 \mathrm{\; AU}$) for the properties of the solar birth cluster. We find that the existence of these orbits, if they were in place in the Sun's birth cluster phase, would limit the product of the stellar number density and the Sun's residence time in the birth cluster to $\lesssim 5 \times 10^3 \mathrm{\; Myr \; pc^{-3}}$, as compared to the weaker limit $\lesssim 5 \times 10^4 \mathrm{\; Myr \; pc^{-3}}$ implied by the low inclinations of the cold classical Kuiper belt.

\end{abstract}


\section{Introduction}

The Sun, like the majority of stars, is thought to have been born in a stellar cluster \citep{2003ARA&A..41...57L, 2010ARA&A..48...47A}. One of the strongest lines of evidence is the detection of daughter products of short-lived radionuclides in meteorites \citep{2007ApJ...662.1268O, 2010ApJ...711..597O}, which would suggest that the Sun formed in a stellar cluster in which at least one Wolf-Rayet star was present and/or one core-collapse supernova occurred 
\citep{2010ARA&A..48...47A, 2019A&A...622A..69P, 2023A&A...670A.105A}. In addition, arguments have posited that the gravitational influence of passing stars from the solar birth cluster could reproduce certain properties of known dynamical features of the distant Kuiper Belt such as the detached disk, the scattered disk, and the Oort cloud \citep{2004Natur.432..598K, 2004AJ....128.2564M, 2006Icar..184...59B, 2012Icar..217....1B, 2015MNRAS.446.3788B, 2023Icar..40615738N, 2024NatAs...8.1380P}. Upper limits on the product of the stellar density $n$ and Sun's residence time in the solar birth cluster $\tau$ have been derived through dynamical constraints based on the survival and stability of the outer planets in their current orbits \citep{2001Icar..150..151A, 2007MNRAS.378.1207M, 2011MNRAS.411..859M, 2013A&A...549A..82P, 2022MNRAS.515.5942B, 2015MNRAS.448..344L, 2015PhyS...90f8001P}. The strongest upper limit quoted in the literature on $\chi \equiv n \times \tau$, however, is based on the observed free inclination distribution of the cold classical Kuiper belt \citep{2020AJ....159..101B}. 

There is a population of small bodies in the distant solar system that may offer more constraining power on $\chi$ than the cold classicals; namely, high-perihelion ($q > 40\mbox{\;AU}$), high-semimajor-axis ($a > 400\mbox{\;AU}$) objects, sometimes referred to as `distant sednoids' (for reference, Sedna has $q=506$ AU and $a=76$ AU). All nine such objects\footnote{While the nominal orbit of 2020 MQ53 technically satisfies the criteria on $q$ and $a$, its orbit is extremely uncertain; hence it is excluded from this work. The inclination of 2020 MQ53 is $\sim 70^{\circ}$.} discovered at the time of this work have inclinations of $< 20^{\circ}$.\footnote{If we instead choose $a > 300 \mathrm{\; AU}$, six additional objects join the sample, and all have inclinations of $< 26^{\circ}$. This illustrates that the specific $400 \mathrm{\; AU}$ threshold used in this work is not essential to the conclusions.} The low inclinations of distant sednoids \citep{2025arXiv250516317H} limit the degree of dynamical excitation that the solar system could have plausibly received from passing stars in the solar birth cluster.\footnote{Selection effects do bias against the discovery of high-inclination objects. We investigate this bias in the paper and find that is not sufficient to explain the surprisingly low inclinations of these objects.} The birth cluster is a more likely source of dynamically influential stellar flybys as compared to the current solar neighborhood given its low velocity dispersion and high number density.

Here, we reproduce the experiment by \cite{2020AJ....159..101B} testing the influence of stellar flybys on the cold classical inclination distribution and perform new experiments studying the sensitivity of high-$q$, high-$a$ objects to stellar flybys in the solar birth cluster. We derive new limits on the product of the stellar number density in the solar birth cluster and the Sun's residence time in the cluster. Rather than focusing on the effects of a single flyby, we simulate a suite of randomly generated stellar flyby histories to establish a link between the conditions in the solar birth cluster and the likelihood of disrupting orbits in the outer solar system.

\section{Simulations}

We carry out all $n$-body simulations in this work with the \texttt{AIRBALL} \citep{airball} and \texttt{REBOUND} \citep{2012A&A...537A.128R} packages using the \texttt{IAS15} hybrid symplectic integrator \citep{2015MNRAS.446.1424R}. Each simulation includes the Sun, Neptune, and a set of massless test particles. A set of stellar flybys is then successively applied to the system. The stellar population has a velocity dispersion of $\sigma_{\rm cluster}= 1\mbox{\;km\,s}^{-1}$ and masses sampled between $0.08$ and $100\,M_\odot$ from a piecewise combination of the \cite{1955ApJ...121..161S} initial mass function (IMF) for $M_{\star} \geq 1 \mathrm{\;M_{\odot}}$ and the \cite{2003PASP..115..763C} IMF for $M_{\star} < 1 \mathrm{\;M_{\odot}}$. We only consider flybys with impact parameters less than $b_{\rm max} = 10^4\mbox{\;AU}$, and each flyby begins and ends at a star-Sun distance of $10^5 \mbox{\;AU}$. The \texttt{AIRBALL} package applies randomly oriented flybys drawn from this stellar population using \texttt{REBOUND}. We record the orbital elements of each test particle after each flyby, and any that become unbound or reach a semimajor axis of $> 10^5\mbox{\;AU}$ are removed from the simulation.

Note that aside from Poisson sampling errors and given a velocity dispersion $\sigma_{\rm cluster}$, there exists a one-to-one mapping between the number of flybys within a certain impact parameter, $N(<b_{\rm max})$ and the product of the stellar number density and the Sun's residence time in the birth cluster, $\chi\equiv n\tau$. Specifically,
\begin{equation}
\chi = \frac{\langle N(<b_{\rm max})\rangle}{4 \sqrt{\pi} b_{\rm max}^2 \sigma_{\rm cluster}} \; \; .
\end{equation}

\subsection{Cold Classical Limit}
\label{coldclassicals}

The cold classical Kuiper belt has semimajor axes between 42 and 47 AU, and free inclinations of $\lesssim 4^{\circ}$ \citep{2021ARA&A..59..203G}. Its free inclination distribution is well-described by a Rayleigh distribution with mode $\sigma = 1.7^{\circ}$. In this section, we will simulate the inclination excitation of the cold classical Kuiper belt by stellar flybys. Specifically, we will reproduce and then build on the work of \cite{2020AJ....159..101B} and argue for a more conservative constraint than they give. Then, in the following section, we will investigate the limit on stellar flybys set by inclination excitation of the distant sednoids.

Each cold-classical simulation contains 160 test particles. The initial values of semimajor axis are randomly selected from a uniform distribution in the range $42$--$47\mathrm{\; AU}$, the values of inclination are randomly selected from a Rayleigh distribution with mode $\sigma = 1.7^{\circ}$, and the eccentricities are set to zero. The longitudes of node ($\Omega$) and mean anomalies ($M$) are drawn randomly from uniform distributions in the range $0$--$2\pi$. In addition, Neptune's inclination and eccentricity are set to zero such that there is zero forced component of the test particles' inclinations and eccentricities. We run 192 such simulations. After each stellar flyby we apply a $42$--$47\mbox{\;AU}$ semimajor axis mask to ensure that the selected test particles always correspond to orbits that fall within the cold classical Kuiper belt, and therefore would have been selected in the first place. Unlike the removal condition for particles on unbound or large semimajor axis ($a > 10^5\mathrm{\; AU}$) orbits, the mask does not remove particles from the simulation -- it simply filters them out from the displayed results.

\begin{figure}
 \centering
\includegraphics[width=\linewidth]{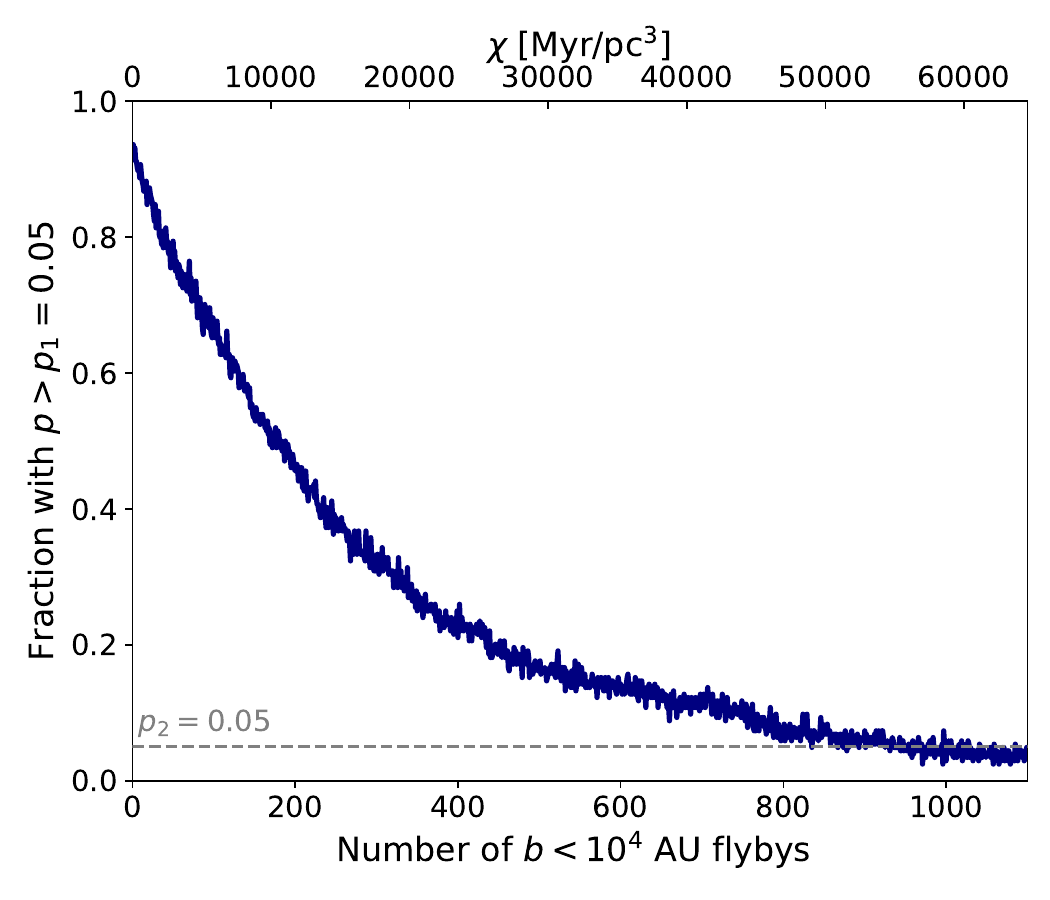}
\caption{Fraction of flyby histories that are compatible with the observed inclination distribution of cold classicals as a function of number of flybys with impact parameter $< 10^4\mathrm{\; AU}$. The initial condition for the inclination distribution is a Rayleigh distribution with a mode of $1.7^{\circ}$. The resulting upper limit on $\chi$, if we adopt a KS test threshold of $p_1 = 0.05$ and require at least $p_2 = 0.05$ of all simulations to exceed this threshold in order to remain compatible, is $\lesssim 6 \times 10^4 \mathrm{\;Myr\; pc^{-3}}$. The fraction starts near 1 because the KS test is relative to an assumed distribution of inclinations that matches the initial distribution.}
\label{fig:coldclassicals}
\end{figure}

As in \cite{2020AJ....159..101B}, we then perform a Kolmogorov–Smirnov (KS) test at each timestep between the inclinations of the test particles passing the mask and a Rayleigh distribution with $\sigma = 1.7^{\circ}$ to match the observed inclination distribution of the cold classicals. In order to derive a limit on $\chi$ from the results, one needs to choose two statistical thresholds: one $p$-value threshold, $p_1$, that defines compatibility based on the result of the KS test for a particular simulation after a particular number of flybys, and one fractional threshold $p_2$ that defines the proportion of compatible $p_1$-values below which we rule out a particular value of number of flybys. Thus, for example, \cite{2020AJ....159..101B} define the simulated distribution of inclinations to be incompatible with the assumed Rayleigh distribution with $\sigma=1.7^\circ$ if the KS test implies that the distributions are different with $99.865\%$ ($3\sigma$) confidence ($p_1=0.00135$), and that a particular number of flybys is ruled out if less than $p_2=75\%$ of the simulations yield compatible KS tests. We verify that given the choices made for these values in \cite{2020AJ....159..101B}, the limit on $\chi$ found in our simulations is indeed the limit quoted in that paper, namely $\chi= 2 - 3 \times 10^4 \mathrm{\; Myr \; pc^{-3}}$. However, we believe that the choice $p_2=0.75$ is too large to represent a robust limit. In Figure \ref{fig:coldclassicals}, we plot the evolution of the compatible fraction for $p_1 = 0.05$ and show that $95\%$ of flyby histories excite the inclinations of the cold classicals such that they are incompatible with the initial condition ($p_2 = 0.05$) for 
$\chi \gtrsim 6 \times 10^4 \mathrm{\; Myr \; pc^{-3}}$.\footnote{If we instead use $p_1 = 0.00135$ (the choice made by \citealt{2020AJ....159..101B}) alongside $p_2 = 0.05$, the limit becomes $\chi \sim 8 \times 10^4 \mathrm{\; Myr \; pc^{-3}}$.} Adopting $p_2 = 0.05$ instead of $p_2 = 0.75$ excludes $\chi$ only when a large fraction ($95\%$) of flyby histories are incompatible, as opposed to the case of $75\%$ of flyby histories still remaining viable. We believe this is a more robust limit because $p_2$ is near zero, and we use the same thresholds for direct comparison when deriving our new limit in Section \ref{newlimit}. 

\subsection{Distant Sednoid Limit}
\label{newlimit}

At the time that these simulations were set up, there were nine known TNOs with $q > 40\mbox{\;AU}$ and $a > 400\mbox{\;AU}$. All of the nine TNOs have remarkably low inclinations of $< 20^{\circ}$.\footnote{Since then, one new such TNO with $q > 40\mbox{\;AU}$ and $a > 400\mbox{\;AU}$ has been discovered, namely 2017 OF201 \citep{2025arXiv250515806C}. Its inclination is also $< 20^{\circ}$ so adding it to the sample here would only strengthen the derived limits on $\chi$.} 

\begin{figure}
 \centering
\includegraphics[width=\linewidth]{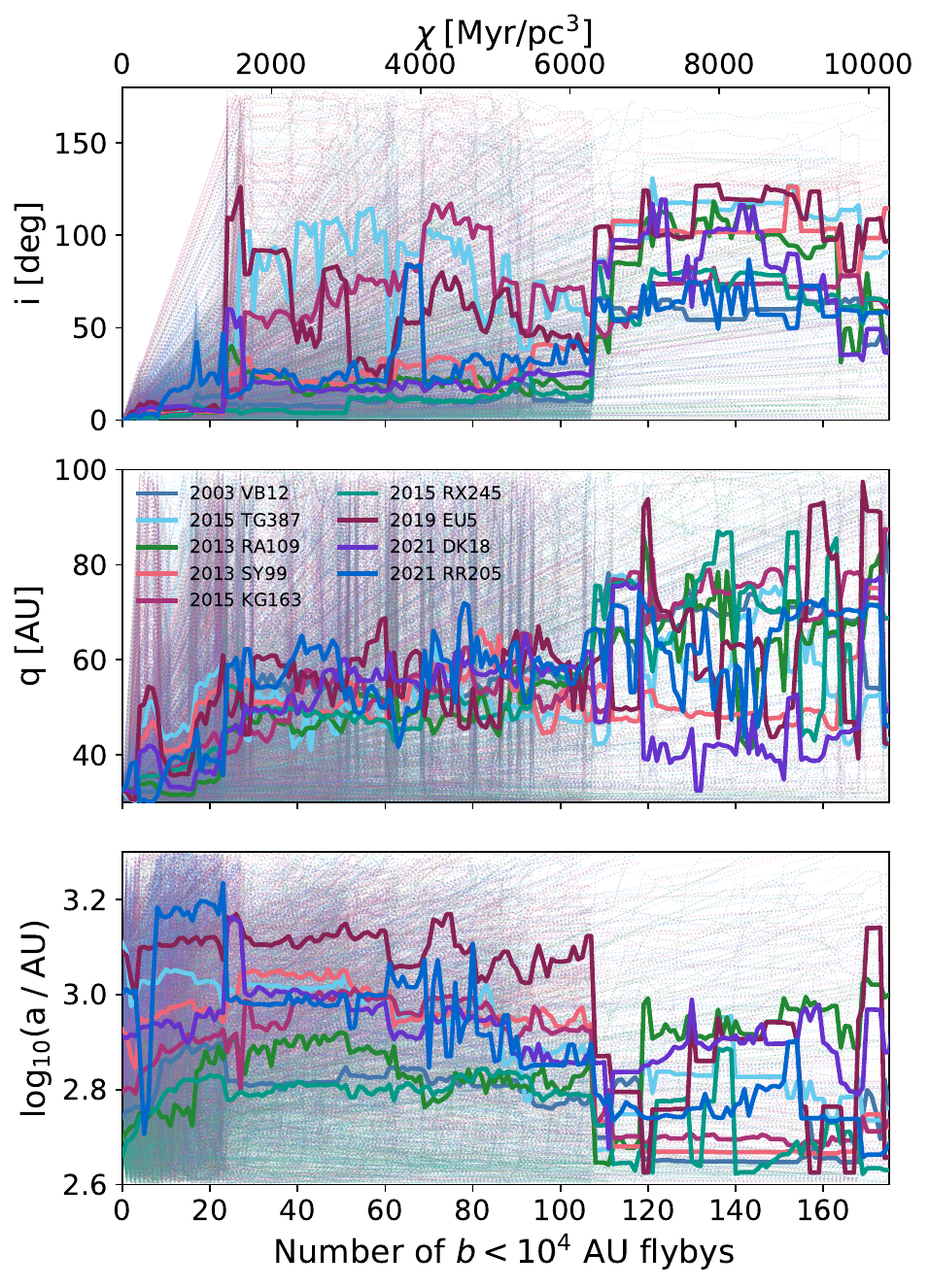}
\caption{Example of inclination, perihelion, and semimajor axis evolution for each of the nine TNOs (median values in bold, individual clones in the background) in an individual simulation. Clones are only plotted and taken into the median calculation if they instantaneously pass the mask described in the text.}
\label{fig:9TNO_example}
\end{figure}

Each simulation contains 2880 test particles (320 clones corresponding to each of the nine TNOs). So that we can study the detachment of perihelia and the heating of inclinations -- and set a conservative limit on $\chi$ -- each clone of each TNO is started with an inclination of zero, a perihelion value drawn randomly from a uniform distribution in the range $30$--$35\mbox{\;AU}$, a mean anomaly drawn randomly from a uniform distribution in the range $0$--$2\pi$, and the other orbital elements left unchanged. We run 219 such simulations, each corresponding to a different history of stellar flybys. After each simulation run, we apply a $30$--$100\mbox{\;AU}$ perihelion mask for the test particles to qualify as observable TNOs and a semimajor axis mask of $400 - 2000\mbox{\;AU}$ so that the selected test particles correspond to orbits that we would have selected in the first place given the original selection criterion on the low end and the furthest known such TNOs on the high end. Figure \ref{fig:9TNO_example} illustrates an example of the orbital element evolution that the clones of the nine TNOs experience in an individual simulation.

\begin{figure}
 \centering
\includegraphics[width=\linewidth]{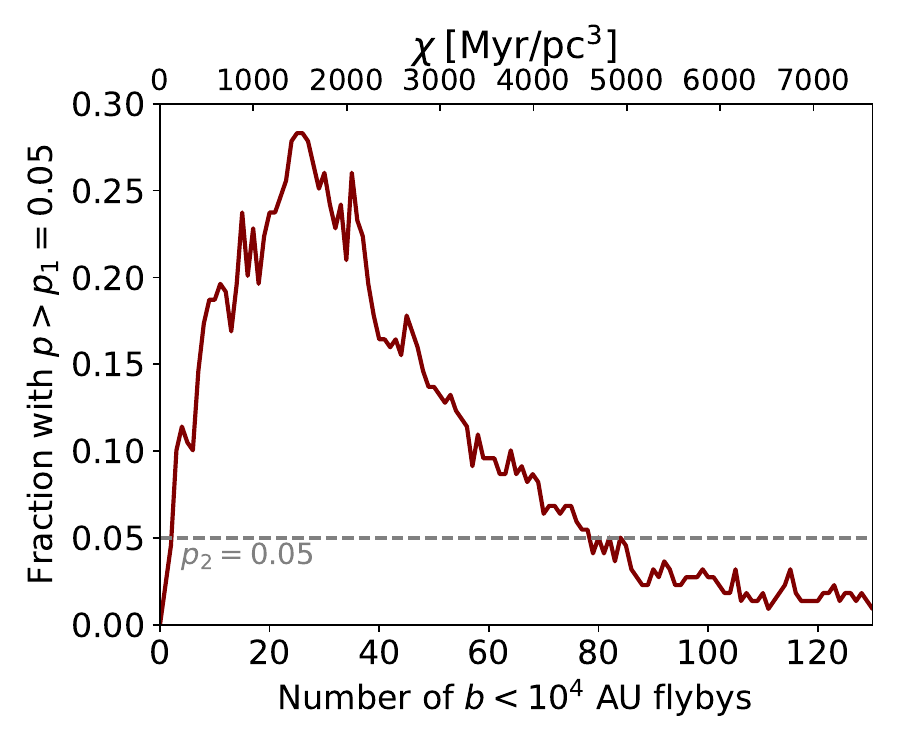}
\caption{Fraction of flyby histories that are compatible with the observed inclination distribution of distant sednoids (initial inclinations are set to zero) as a function of number of flybys with impact parameter $< 10^4\mathrm{\; AU}$. The resulting upper limit on $\chi$, if we adopt a KS test threshold of $p_1 = 0.05$ and require at least $p_2 = 0.05$ of all simulations to exceed this threshold in order to remain compatible, is $\lesssim 5 \times 10^3 \mathrm{\;Myr\; pc^{-3}}$. The fraction starts near zero because the initial inclinations are set at zero.}
\label{fig:9TNO_limit}
\end{figure}

We then perform a Kolmogorov–Smirnov (KS) test at each timestep between the inclinations of the test particles passing the mask, weighted\footnote{If the inclinations are weighted equally, the constraint on $\chi$ strengthens by a factor of two.} by a factor of $(1/\sin{i})$ with a maximum weight at $i = 1^{\circ}$, and the actual inclinations of the nine TNOs. This weighting acts as an approximate correction for the bias of outer solar system surveys towards low-inclination objects, given the usual focus near the ecliptic plane. In Figure \ref{fig:9TNO_limit}, we plot the evolution of the compatible fraction for $p_1 = 0.05$ and show that $95\%$ of flyby histories excite the inclinations of the distant sednoids from zero such that they are incompatible with the initial condition ($p_2 = 0.05$) by $\chi \sim 5 \times 10^3 \mathrm{\; Myr \; pc^{-3}}$.\footnote{If we instead use $p_1 = 0.00135$ (the choice made by \citealt{2020AJ....159..101B}) alongside $p_2 = 0.05$, the limit becomes $\chi \sim 8 \times 10^3 \mathrm{\; Myr \; pc^{-3}}$.} These are the same statistical thresholds used in Section \ref{coldclassicals}, yet the limit is an order of magnitude better primarily because of the low orbital binding energies of this population relative to the cold classicals. Also, this limit is more conservative than the cold classicals one because here the inclinations are started at zero rather than their observed values. Furthermore, we note that $\chi \sim 1.5 \times 10^3 \mathrm{\; Myr \; pc^{-3}}$ maximizes the fraction of flyby histories compatible with the observed inclination distribution, implying that if the distant sednoids did not gain any inclination when they were scattered in semimajor axis (a conservative assumption), the best-fit to the inclination heating required by stellar flybys corresponds to $\chi \sim 1.5 \times 10^3 \mathrm{\; Myr \; pc^{-3}}$ -- a factor of three lower than the constraint. 

We may also check whether stellar flybys in the birth cluster can detach the perihelia of the distant sednoids without excessive excitation of their inclinations. To do this, we run a KS test comparing perihelion values of the particles passing the mask to the observed perihelion values of the nine distant sednoids. Because we started the test particles with $q$ in the range $30 - 35\mbox{\;AU}$, initially all simulations are incompatible with the observations, but after only a couple of flybys ($\chi \sim 10^2 \mathrm{\; Myr / pc^3}$), the perihelia are detached to a level that $> 5\%$ of simulations are compatible with the observed distribution, and by $\chi \sim 10^3 \mathrm{\; Myr / pc^3}$ this fraction is $\sim 50\%$. Because the detection probability falls off steeply with perihelion distance, we are unable to place an upper limit on $\chi$ based on the observed perihelion distribution of distant sednoids -- we can only say that a large $\chi$ is not required to detach their perihelia to match observations. Closer-in ($a < 400 \mathrm{\; AU}$) sednoids require a greater number of flybys to detach their perihelia from Neptune.

\section{Discussion}
Using the inclination distribution of the distant sednoids, we derived a limit on the product of the stellar number density of and the Sun's residence time in the solar birth cluster of $\lesssim 5 \times 10^3 \mathrm{\;Myr\; pc^{-3}}$. This limit is stronger by an order of magnitude than the limit derived from the inclination distribution of the cold classicals using the same values for $p_1$ and $p_2$ as used here. The strength comes from the sensitivity of these distant objects' orbits to perturbations from passing stars, and the principal limitation comes from the small number of known distant sednoids.

The number of stars in the solar birth cluster necessary for the solar system to have experienced an episode of short-lived radionuclide enrichment is somewhat uncertain, with estimates ranging from $5 \times 10^2 - 2 \times 10^4$ \citep{2023A&A...670A.105A}. The relation between cluster size and membership is also uncertain, and whether mean number density increases or decreases with membership depends on the environment. A typical value for the mean number density of a stellar cluster is $\sim 10^2\mathrm{\; pc^{-3}}$ \citep{2010ARA&A..48...47A}; our upper limit of $\chi \sim 5 \times 10^3 \mathrm{\; Myr \; pc^{-3}}$ would then translate to an upper limit of $\sim 50 \mathrm{\; Myr}$ for the Sun's residence time in the birth cluster. Of course, the quantity $\chi$ actually involves an integral $\int n(t)\;dt$, implying that the upper limit on residence time depends on the time-evolution of $n$.

While our limit makes the conservative assumption that the inclinations of distant sednoids start at zero, it also only applies if these objects were scattered to their current semimajor axes early on in the Sun's birth cluster residency. A similar caveat holds for the cold classical limit \citep{2020AJ....159..101B}; in general any solar system objects used to place a limit on $\chi$ need to be placed on their current orbits on a timescale much shorter than the Sun's cluster residence time. The cold classical Kuiper belt's location within the protoplanetary disk, and its low eccentricities and inclinations, are consistent with in‑situ formation in the gaseous protoplanetary disk. Given the short lifetime of the gaseous disk, these objects were plausibly in place well before the Sun left its birth environment. In contrast, the distant sednoids have perihelia beyond Neptune and semimajor axes far exceeding typical disk radii \citep{2010ARA&A..48...47A}, so they likely did not form in place. The mechanism of formation of the distant sednoids is not well-understood. Our sednoid‑based constraint should therefore be interpreted as conditional: it is the stronger bound if the distant sednoids were emplaced early in the Sun's birth cluster phase, but it does not apply if they were emplaced after the Sun left its birth cluster.

The finding here that the distant sednoids provide a limit on $\chi$ an order of magnitude stronger than the cold classicals is consistent with the following analytic estimate. In the regime of interest, Equation (43) of \cite{2020AJ....159..101B} implies $\chi \propto \Delta i \: a^{-3/2}$, where $\Delta i$ is the inclination kick magnitude and $a$ is the semimajor axis of the population. If one adopts $\Delta i$ of $1.7^{\circ}$ and $20^{\circ}$ to disrupt the two populations (given their observed inclinations), and typical semimajor axes of $45 \mathrm{\; AU}$ and $10^3 \mathrm{\; AU}$, respectively, the analytic expectation is consistent with the numerical result: the distant sednoids should provide a constraint on $\chi$ that is an order of magnitude stronger than the cold classicals.

The Legacy Survey of Space and Time (LSST) on the Vera C. Rubin Observatory is expected to increase the number of known TNOs by an order of magnitude \citep{2025AJ....170...99K}. Finding more distant sednoids will reduce the uncertainty on the limit derived here.

\section*{Acknowledgements}
We are pleased to acknowledge that the work reported in this Letter was substantially performed using the Princeton Research Computing resources at Princeton University, which is a consortium of groups led by the Princeton Institute for Computational Science and Engineering (PICSciE) and Office of Information Technology's Research Computing.


\bibliography{bib}{}

\begin{thebibliography}{}
\expandafter\ifx\csname natexlab\endcsname\relax\def\natexlab#1{#1}\fi
\providecommand{\url}[1]{\href{#1}{#1}}

\bibitem[{{Adams}(2010)}]{2010ARA&A..48...47A}
{Adams}, F.~C. 2010, \araa, 48, 47

\bibitem[{{Adams} \& {Laughlin}(2001)}]{2001Icar..150..151A}
{Adams}, F.~C., \& {Laughlin}, G. 2001, \icarus, 150, 151

\bibitem[{{Arakawa} \& {Kokubo}(2023)}]{2023A&A...670A.105A}
{Arakawa}, S., \& {Kokubo}, E. 2023, \aap, 670, A105

\bibitem[{{Batygin} {et~al.}(2020){Batygin}, {Adams}, {Batygin}, \& {Petigura}}]{2020AJ....159..101B}
{Batygin}, K., {Adams}, F.~C., {Batygin}, Y.~K., \& {Petigura}, E.~A. 2020, \aj, 159, 101

\bibitem[{{Brasser} {et~al.}(2006){Brasser}, {Duncan}, \& {Levison}}]{2006Icar..184...59B}
{Brasser}, R., {Duncan}, M.~J., \& {Levison}, H.~F. 2006, \icarus, 184, 59

\bibitem[{{Brasser} {et~al.}(2012){Brasser}, {Duncan}, {Levison}, {Schwamb}, \& {Brown}}]{2012Icar..217....1B}
{Brasser}, R., {Duncan}, M.~J., {Levison}, H.~F., {Schwamb}, M.~E., \& {Brown}, M.~E. 2012, \icarus, 217, 1

\bibitem[{{Brasser} \& {Schwamb}(2015)}]{2015MNRAS.446.3788B}
{Brasser}, R., \& {Schwamb}, M.~E. 2015, \mnras, 446, 3788

\bibitem[{{Brown} \& {Rein}(2022)}]{2022MNRAS.515.5942B}
{Brown}, G., \& {Rein}, H. 2022, \mnras, 515, 5942

\bibitem[{{Brown} {et~al.}(2024){Brown}, {Rein}, {Mohsin}, {Chao-Ming Lam}, {Generozov}, {He}, \& {Shi}}]{airball}
{Brown}, G., {Rein}, H., {Mohsin}, H., {et~al.} 2024.
\newblock \url{https://airball.readthedocs.io/}

\bibitem[{{Chabrier}(2003)}]{2003PASP..115..763C}
{Chabrier}, G. 2003, \pasp, 115, 763

\bibitem[{{Cheng} {et~al.}(2025){Cheng}, {Li}, \& {Yang}}]{2025arXiv250515806C}
{Cheng}, S., {Li}, J., \& {Yang}, E. 2025, arXiv e-prints, arXiv:2505.15806

\bibitem[{{Gladman} \& {Volk}(2021)}]{2021ARA&A..59..203G}
{Gladman}, B., \& {Volk}, K. 2021, \araa, 59, 203

\bibitem[{{Hu} {et~al.}(2025){Hu}, {Huang}, {Gladman}, \& {Zhu}}]{2025arXiv250516317H}
{Hu}, Q., {Huang}, Y., {Gladman}, B., \& {Zhu}, W. 2025, arXiv e-prints, arXiv:2505.16317

\bibitem[{{Kenyon} \& {Bromley}(2004)}]{2004Natur.432..598K}
{Kenyon}, S.~J., \& {Bromley}, B.~C. 2004, \nat, 432, 598

\bibitem[{{Kurlander} {et~al.}(2025){Kurlander}, {Bernardinelli}, {Schwamb}, {Juri{\'c}}, {Murtagh}, {Chandler}, {Merritt}, {Nesvorn{\'y}}, {Vokrouhlick{\'y}}, {Jones}, {Fedorets}, {Cornwall}, {Holman}, {Eggl}, {Oldag}, {West}, {Kubica}, {Yoachim}, {Moeyens}, {Kiker}, \& {Buchanan}}]{2025AJ....170...99K}
{Kurlander}, J.~A., {Bernardinelli}, P.~H., {Schwamb}, M.~E., {et~al.} 2025, \aj, 170, 99

\bibitem[{{Lada} \& {Lada}(2003)}]{2003ARA&A..41...57L}
{Lada}, C.~J., \& {Lada}, E.~A. 2003, \araa, 41, 57

\bibitem[{{Li} \& {Adams}(2015)}]{2015MNRAS.448..344L}
{Li}, G., \& {Adams}, F.~C. 2015, \mnras, 448, 344

\bibitem[{{Malmberg} {et~al.}(2011){Malmberg}, {Davies}, \& {Heggie}}]{2011MNRAS.411..859M}
{Malmberg}, D., {Davies}, M.~B., \& {Heggie}, D.~C. 2011, \mnras, 411, 859

\bibitem[{{Malmberg} {et~al.}(2007){Malmberg}, {de Angeli}, {Davies}, {Church}, {Mackey}, \& {Wilkinson}}]{2007MNRAS.378.1207M}
{Malmberg}, D., {de Angeli}, F., {Davies}, M.~B., {et~al.} 2007, \mnras, 378, 1207

\bibitem[{{Morbidelli} \& {Levison}(2004)}]{2004AJ....128.2564M}
{Morbidelli}, A., \& {Levison}, H.~F. 2004, \aj, 128, 2564

\bibitem[{{Nesvorn{\'y}} {et~al.}(2023){Nesvorn{\'y}}, {Bernardinelli}, {Vokrouhlick{\'y}}, \& {Batygin}}]{2023Icar..40615738N}
{Nesvorn{\'y}}, D., {Bernardinelli}, P., {Vokrouhlick{\'y}}, D., \& {Batygin}, K. 2023, \icarus, 406, 115738

\bibitem[{{Ouellette} {et~al.}(2007){Ouellette}, {Desch}, \& {Hester}}]{2007ApJ...662.1268O}
{Ouellette}, N., {Desch}, S.~J., \& {Hester}, J.~J. 2007, \apj, 662, 1268

\bibitem[{{Ouellette} {et~al.}(2010){Ouellette}, {Desch}, \& {Hester}}]{2010ApJ...711..597O}
---. 2010, \apj, 711, 597

\bibitem[{{Pfalzner}(2013)}]{2013A&A...549A..82P}
{Pfalzner}, S. 2013, \aap, 549, A82

\bibitem[{{Pfalzner} {et~al.}(2024){Pfalzner}, {Govind}, \& {Portegies Zwart}}]{2024NatAs...8.1380P}
{Pfalzner}, S., {Govind}, A., \& {Portegies Zwart}, S. 2024, Nature Astronomy, 8, 1380

\bibitem[{{Pfalzner} {et~al.}(2015){Pfalzner}, {Davies}, {Gounelle}, {Johansen}, {M{\"u}nker}, {Lacerda}, {Portegies Zwart}, {Testi}, {Trieloff}, \& {Veras}}]{2015PhyS...90f8001P}
{Pfalzner}, S., {Davies}, M.~B., {Gounelle}, M., {et~al.} 2015, \physscr, 90, 068001

\bibitem[{{Portegies Zwart}(2019)}]{2019A&A...622A..69P}
{Portegies Zwart}, S. 2019, \aap, 622, A69

\bibitem[{{Rein} \& {Liu}(2012)}]{2012A&A...537A.128R}
{Rein}, H., \& {Liu}, S.~F. 2012, \aap, 537, A128

\bibitem[{{Rein} \& {Spiegel}(2015)}]{2015MNRAS.446.1424R}
{Rein}, H., \& {Spiegel}, D.~S. 2015, \mnras, 446, 1424

\bibitem[{{Salpeter}(1955)}]{1955ApJ...121..161S}
{Salpeter}, E.~E. 1955, \apj, 121, 161

\end{thebibliography}
\bibliographystyle{aasjournal}

\end{document}